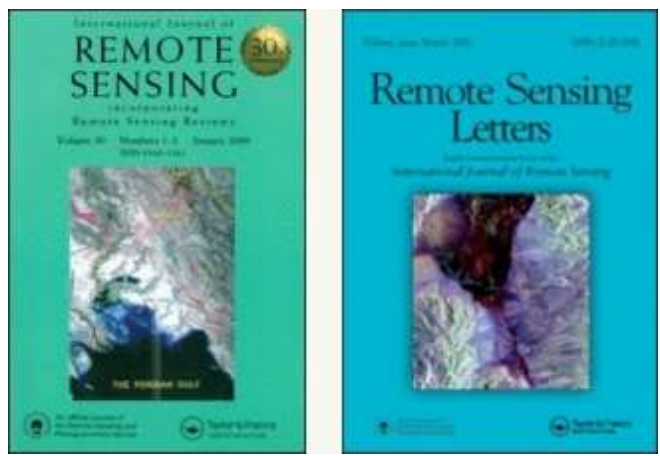

# Segment-based fusion of multi-sensor multi-scale satellite soil moisture retrievals

# Segment-based fusion of multi-sensor multi-scale satellite soil moisture retrievals

Reza Attarzadeh[a]*, Hossein Bagheri[b], Iman Khosravi[b], Saeid Niazmardi[c], Davood Akbari[d]

[a] Department of Geodesy and Geomatics Engineering, South Tehran Branch, Islamic Azad University, Tehran, Iran

[b] Department of Geomatics Engineering, Faculty of Civil Engineering & Transportation, University of Isfahan, Isfahan, Iran

[c] Remote Sensing Department, Faculty of Civil & Surveying Engineering, Kerman Graduate University of Advanced Technology (KGUT), Kerman, Iran

[d] Department of Geomatic Engineering, Faculty of Engineering, University of Zabol, Zabol, Iran

# Segment-based fusion of multi-sensor multi-scale satellite soil moisture retrievals

Synergetic use of sensors for soil moisture retrieval is attracting considerable interest due to the different advantages of different sensors. Active, passive, and optic data integration could be a comprehensive solution for exploiting the advantages of different sensors aimed at preparing soil moisture maps. Typically, pixel-based methods are used for multi-sensor fusion. Since, different applications need different scales of soil moisture maps, pixel-based approaches are limited for this purpose. Object-based image analysis employing an image object instead of a pixel could help us to meet this need. This paper proposes a segment-based image fusion framework to evaluate the possibility of preparing a multi-scale soil moisture map through integrated Sentinel-1, Sentinel-2, and Soil Moisture Active Passive (SMAP) data. The results confirmed that the proposed methodology was able to improve soil moisture estimation in different scales up to 20% better compared to pixel-based fusion approach.

## 1. Introduction

Soil moisture is an important meteorological and climate parameter for diverse applications such as drought monitoring, irrigation planning (Bazzi et al., 2019), soil-vegetation-atmospheric interface process (Koster et al., 2004), climate investigations, etc. Typically, soil moisture measurements are provided by in-situ probes. However, in-situ measuring of soil moisture is an expensive and tedious task that can be applied to small areas. Alternatively, satellite sensors can provide an opportunity to retrieve soil moisture in large area coverage. Satellite-based soil moisture estimation is generally realized by using measurements in the optical and microwave ranges of the electromagnetic spectrum. In the microwave range, a wide variety of sensors in different modes; passive microwave sensors (Jackson et al., 2010), radar scatterometer or Synthetic Aperture Radar (SAR) (Wagner et al., 2013), and Global Navigation Satellite Systems reflectometry (GNSS-R) (Motte et al., 2016) have been used for soil moisture estimation.

Several satellite sensors with miscellaneous advantageous characteristics and defects have been launched for this purpose. In the literature, the multi-sensor data analysis was considered for preparing accurate soil moisture maps (e.g. Karthikeyan et al. 2017; Amazirh et al. 2018; He et al. 2018; Sabaghy et al. 2018; Bai et al. 2019; Das et al. 2019; Jagdhubera et al. 2019). For example, Soil Moisture Active Passive (SMAP) as a well-known microwave sensor has been launched for soil moisture retrieval by active and passive instruments. With the failure of the radar section of the SMAP mission, its substitution with Sentinel-1A/Sentinel-1B SAR data is offered as a feasible approach for generating high spatial resolution soil moisture products due to its similar orbit configuration. Additionally, optical data collected by Sentinel-2 could also be fused with SMAP and Sentinel-1 observations to retrieve soil moisture by reducing the problems associated with the original products. As a result, high spatial resolution Sentinel data along with highly accurate soil moisture map products of the SMAP mission could lead us to an accurate high spatial resolution soil moisture map (Attarzadeh & Amini, 2019; Bauer-Marschallinger et al. 2019). In fact, a fusion of soil moisture retrievals from various sensors could finally generate more accurate and complementary products (Sabaghy et al. 2017; Amazirh et al. 2018; He et al. 2018; Bai et al. 2019).

On the other hand, in a variety of applications, soil moisture maps with different scales required. In this regard, exploiting image objects instead of pixels for soil moisture retrieval enables us to have more flexibility in generating diverse scales of soil moisture maps based on user needs. Additionally, since each sensor estimates soil moisture at different spatial resolution (scales), providing a multi-scale fusion method (based on the scale) will lead to a better result (Attarzadeh et al. 2018). Thus, Object-Based Image Analysis (OBIA) can be potentially applied for multi-scale soil moisture estimation.

OBIA uses a group of homogeneous pixels in the image instead of using a single pixel, and lead to a segmentation at different scales. Consequently, compatible adjacent pixels are used in soil moisture estimation that finally improves the estimation accuracy.

Until now, only a few investigations have addressed the issue of preparing multi-scale soil moisture maps. Another study estimated the soil moisture from Sentinel-1 and Sentinel-2 by combining pixel- and object-based analyses through a support vector regression (SVR) algorithm (Attarzadeh et al. 2018). In a recent study, the soil moisture retrieval from SMAP at 3 km spatial resolution was fused with high spatial resolution products retrieved from Sentinel-1 and 2 (Attarzadeh & Amini, 2015).

This paper is an extension of the later work mentioned earlier. This study investigates the possibility of fusing low spatial resolution soil moisture retrieval from SMAP with high spatial resolution products generated by Sentinel-1 and 2. For this purpose, an OBIA framework is designed and implemented to fuse soil moisture products at variant scales, spatial resolutions, and coverages. This can be realized by multi-scale, segment-based processing through the proposed framework.

## 2. Materials and methodology

### 2.1. Study area and data sets

The study was carried out in Narok County located in the southwest of Kenya (Figure 1). The district has five agro-climatic zones namely humid, sub-humid, semi-humid, arid and semi-arid. Two-thirds of the district is classified as semi-arid. The agro-ecological zones found in the district include: Tropical Alpine, Upper Highland zones, Lower Highland zones and Medium Height zones. The district experiences bi-modal pattern of rainfall with long rains (Mid March – June) and short rains (September-November). The amount of rainfall is influenced by bi-annual passage of Inter-Tropical Convergence Zone (ITCZ). Rainfall distribution is uneven with high potential areas receiving the highest amount of rainfall ranging from 1200mm – 1800mm p.a. while the lower and drier areas classified as semi-arid receiving 500mm or less p.a. The district experiences a wide variation of temperatures throughout the year with mean annual temperatures varying from 10ºC in Mau escarpment to about 20ºC in the lower drier areas. Soil types in the district are determined by characteristics of the underlying basement rock and range from those developed on mountains to those developed on plains and swamps. The main soil types in the district include: Mollic andosols, luvisols, chromic luvisols, luvic and ando-luvic, phaeozems, chromic vertisols and chromic aerosols.

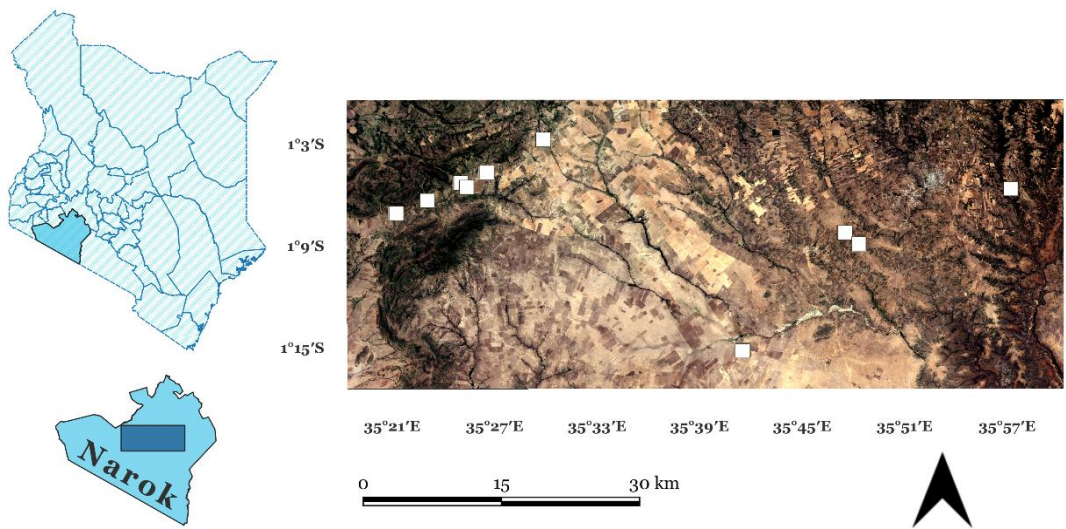

Figure 1. The study area in Kenya (Narok county) with the geographical locations of in situ measurements (The white squares mark the locations of the in situ measurements).

The bases for the soil moisture analysis were Sentinel-1 and Sentinel-2 images of Narok county with a complex topography. The Sentinel-1A image was acquired on September 05, 2016, in Interferometric Wide (IW) swath mode (a 250 km swath width at 5 m by 20 m spatial resolution) with incidence angles ranging from 29º at near range to 46◦ at far range using VV polarization. And the Sentinel-2A optical image was also obtained on August 25, 2016, as close as possible to the radar data considering the cloud coverage. Also, 9 km SMAP L3_SM_P_E descending data (Version 2) for 5 September 2016 was downloaded from NASA's Earth Observing System Data and Information System (EOSDIS). Simultaneous with the Sentinel-1A acquisitions, a total of 35 field measurements were collected. For each sample, four measurements of volumetric soil moisture were conducted in the top 5 cm of soil by means of a calibrated TDR (Time Domain Reflectometry) probe. The minimum and maximum values for in-situ soil moisture measurements were 3.2 and 33.5 vol% respectively. Most of the in-situ data were measured within the cropland, shrubland, and grassland fields. Four types of soil coverage, including bare soil, soil with sparse, moderate, and dense vegetation cover were considered. The canopy height included five categories, namely less than 10 cm, 10–25 cm, 25–50 cm, 50–75 cm, and greater than 75 cm.

*2.2. Pixel-based soil moisture map*

Figure 1 confirms the presence of vegetation cover as one of the land-cover types in our study area. Thus, the Sentinel-2 data has been exploited to disentangle the vegetation effect in the soil moisture retrieval process. In the initial stage of the process, a broad range of radar and optical features have been extracted from Sentinel-1 and Sentinel-2 data. The texture features derived from the VV polarization were also exploited to broaden the radar feature space since only single VV polarization of Sentinel-1 data was available for the study area. The eight features that have been extracted from the grey level co-occurrence matrix (GLCM) matrix and used in this research were contrast, correlation, dissimilarity, entropy, homogeneity, mean, standard deviation, and angular second moment. The most common strategy for disentangle the ambiguity that the vegetation introduces into microwave signals is to utilize optical indices such as Normalized Difference Vegetation Index (NDVI) and Leaf Area Index (LAI). In this study, we attempted using a broader range of optical features to assess the potential usefulness of other optical indices to enhance the Soil Moisture Content (SMC) retrieval process' accuracy. In this context, radiometric indices for three categories—soil, vegetation, and water—were derived from Sentinel-2 data (4, 23, and 2 indicators for the soil, vegetation, and water category, respectively). Then, the most related features with soil moisture variable have been selected using Random Forest-Recursive Feature Elimination (RF- RFE) method. The RF- RFE is a recursive methodology that utilizes data extracted from an Random Forest (RF) model to select (and delete) the less important features from the initial dataset at each iteration. Consequently, six features were selected from of the 37 initial features, including Brightness Index (BI), Green Normalized Difference Vegetation Index (GNDVI), VV polarized backscattering values, second Brightness Index (BI2), GLCM Mean, and Weighted Difference Vegetation Index (WDVI). Given the small training sample size in our study, the SVR method with the Radial Basis Function (RBF) kernel has been used to estimate soil moisture values. Figure 2a presents the output pixel-based soil moisture map using coupled Sentinel-1 and Sentinel-2 data. This Figure illustrates the limitation of pixel-based method for soil moisture estimation in areas with severe topography. Since the backscattering values in the Sentinel-1 image are affected by geometric errors such as radar shadows, layover, etc., the quality of the generated soil moisture map is under the influence of mentioned errors, subsequently.

Another pixel-based product that has been used in this study is the soil moisture map acquired from the SMAP (Figure 2b). The SMAP mission is a unique mission for providing a global soil moisture map by combining passive (radiometer) and active (radar) data. But unfortunately, with the L-band radar failure in July 2015, only the L-band radiometer is still operating and coarse spatial resolution radiometer-based SM data are available. The 9 km SMAP L3_SM_P_E descending data were used in this study because they have a more similar spatial resolution to the average size of objects in the produced OBIA SMC map.

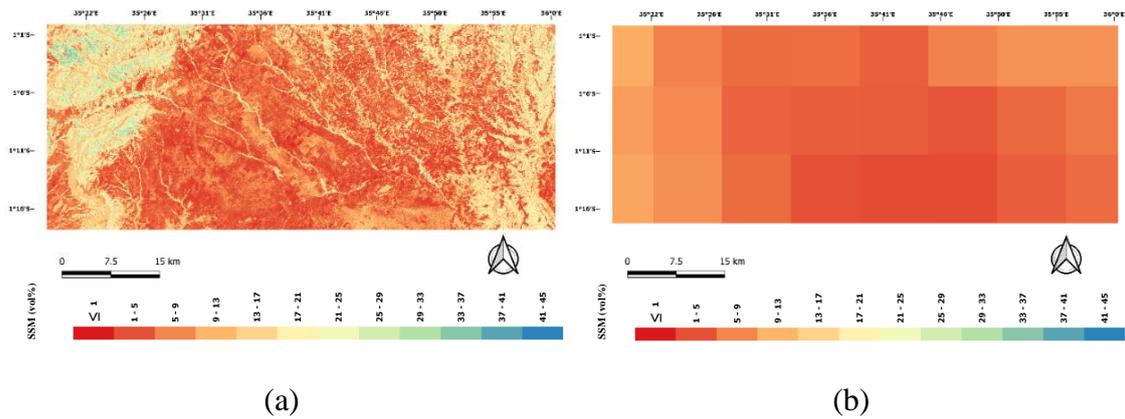

(a)                        (b)

Figure 2. The pixel-based soil moisture map of the study area from: (a) coupled Sentinel-1 and Sentinel-2 data, and (b) SMAP data.

## *2.3. Multi-scale Sentinel-1 and Sentinel-2 soil moisture map*

To answer the major need of different scales of SMC maps for different applications, the OBIA has been exploited. To achieve this aim, in the first step, image objects have been created as a base for preparing multi-scale SMC maps using multi spatial resolution segmentation. The most relevant features chosen in the feature selection are imported into the segmentation process to produce the most relevant objects for the soil moisture variable. It should be noted that the extracted SVR model from the pixel-based approach has been employed to estimate the soil moisture of each image object. For more details on this approach, readers are referred to our previous works (Attarzadeh et al. 2018; Attarzadeh and Amini 2019a; Attarzadeh and Amini 2019b).

Figure 3 demonstrates the soil moisture map prepared from synergetic use of Sentinel-1 and Sentinel-2 data in different scales. Table I also summarizes the statistics of image objects for different scales.

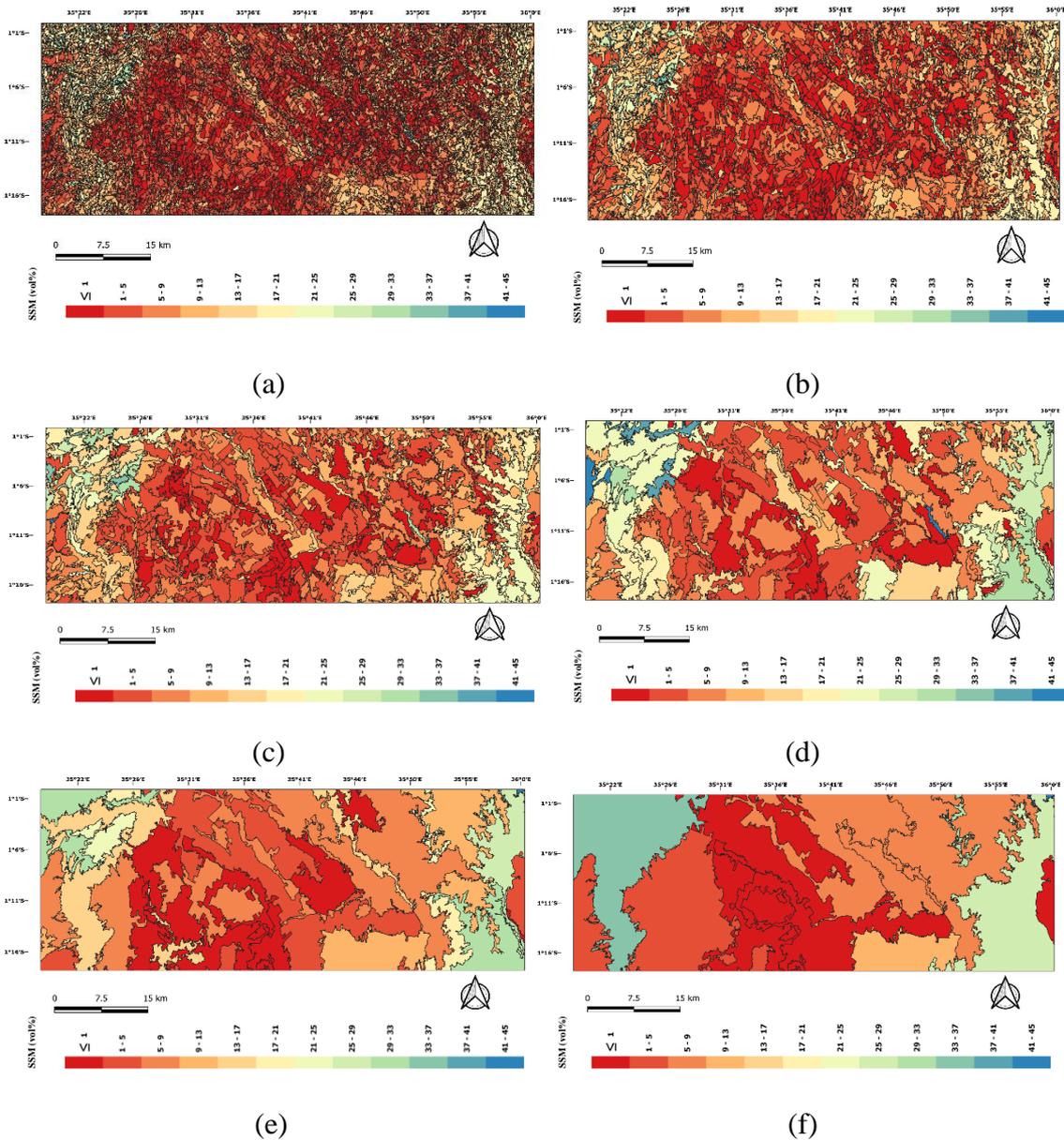

Figure 3. First Scenario: SMC maps with different scale parameters from coupled Sentinel-1 and Sentinel-2 data: (a) SP-64, (b) SP-128, (c) SP-256, (d) SP-512, (e) SP-1024, and (f) SP-2048.

## *2.4. Multi-scale Sentinel-1, Sentinel-2, and SMAP soil moisture map*

Exploiting both characteristics in an OBIA framework with synergetic use of Sentinel and SMAP data seems to better meet the practical application requirements, considering the high spatial resolution and high accuracy of soil moisture retrieval for Sentinel and SMAP data respectively. A pixel with a crisp square shape could not flexibly demonstrate the SM value for a field, especially for small-scale SMC maps. In this research, we assigned the mean values of SMAP soil moisture values within an image object to that object (Figure 4). This means that the shape of image objects comes from Sentinel data while the Soil Moisture (SM) values derive from SMAP data.

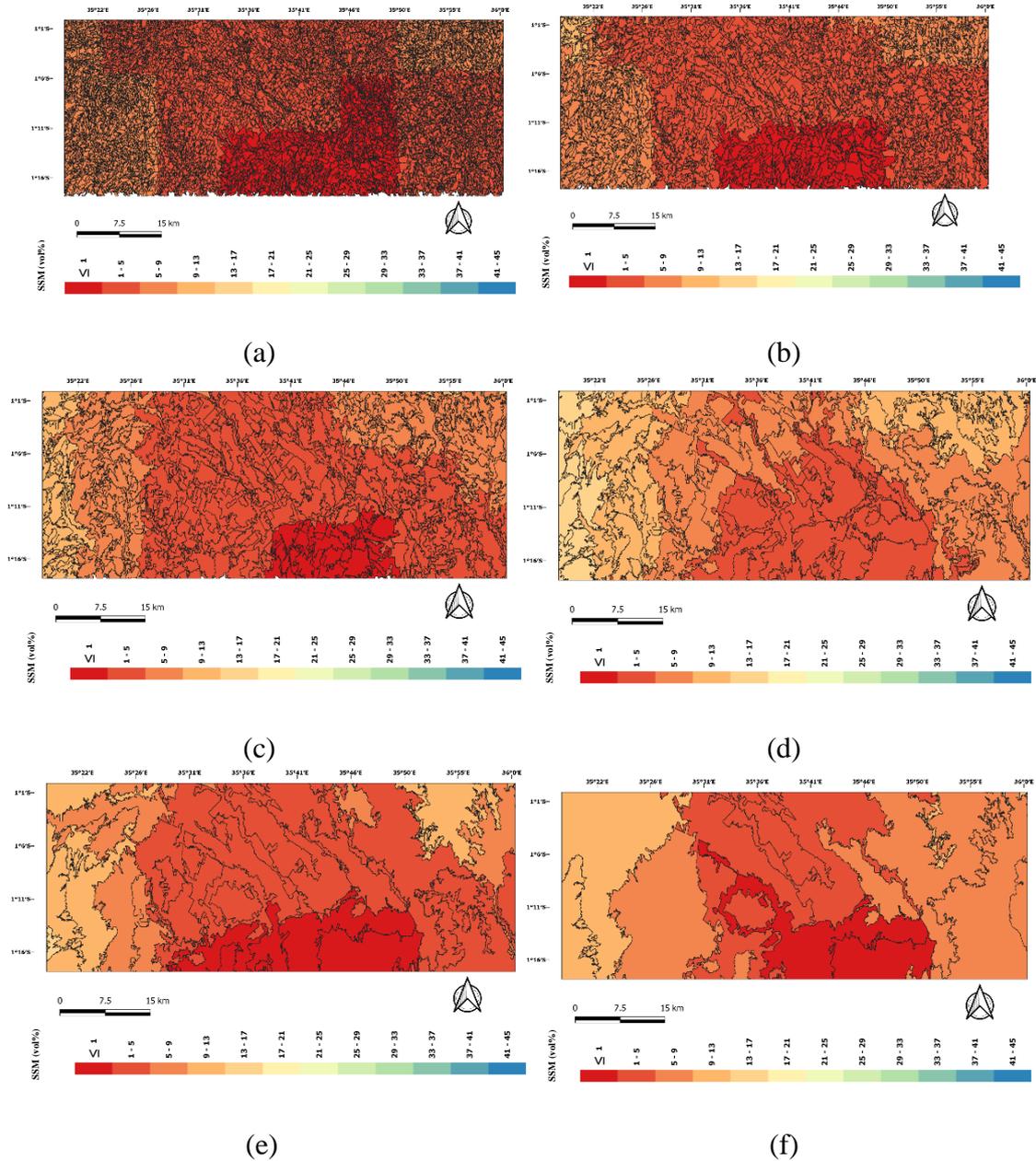

(a) (b) (c) (d) (e) (f)

Figure 4. Second Scenario: SMC maps with different scale parameters from coupled Sentinel-1, Sentinel-2 and SMAP data: (a) SP-64, (b) SP-128, (c) SP-256, (d) SP-512, (e) SP-1024, and (f) SP-2048.

## 3. Results and discussion

Synergetic use of SMAP and Sentinel data through an object-based approach enables us to produce multi-scale SMC maps with acceptable accuracy. This approach allows us to prepare the soil moisture map with the required scale for different applications. As shown in Figure 3 and detailed in Table 1, considering accuracy, the output multi-scale soil moisture map including SMAP data outperforms the output multi-scale soil moisture map without it. As detailed in Table I, the in-situ soil moisture measurements and SMAP soil moisture map mean values are almost equal while the standard deviation of the in-situ soil moisture measurements and Sentinel-1_Sentinel-2 (S1S2) pixel-based soil moisture map are quite equal. Besides, the statistic measures of object-based S1S2 soil moisture map with different scale parameters including mean, standard deviation, RMSE, bias, and coefficient of determination ($R^2$) have heterogeneous behavior than in situ soil moisture measurements. While with the use of SMAP data, the produced SMC map demonstrates similar behavior to the in-situ measurements regarding statistical measures and finally leads to a more homogenous soil moisture map than only using Sentinel data. As shown in Figure 4, the pixel effect has been removed for small-scale SMC maps with 512, 1024, and 2048 scale parameters. This is due to the fact that the average image object area of the SMC map resulting from segmentation with a scale parameter of 512 is equal to 4.45 $km^2$ and is greater than 3 $km^2$ SMAP pixel coverage.

Table 1. Detailed statistics of the image objects for each scale parameter (Column 6: The average area of the image objects in hectare, Column 7: The approximate average printed map scale for corresponding SP)

| SP | NO | NPmi | NPma | NPa | Area (ha) | Map Scale |
|---|---|---|---|---|---|---|
| 64 | 5355 | 4 | 6074 | 724 | 7.24 | 1:270 000 |
| 128 | 1509 | 20 | 22691 | 2720 | 27.20 | 1:521 000 |
| 256 | 428 | 99 | 44397 | 11056 | 110.56 | 1:1051 000 |
| 512 | 126 | 738 | 217983 | 44580 | 445.80 | 1:2111 000 |
| 1024 | 44 | 13288 | 448052 | 153989 | 1539.89 | 1:3924 000 |
| 2048 | 12 | 267091 | 1629081 | 710371 | 7103.71 | 1:8428 000 |

SP: Scale Parameter, NO: number of image objects, NPmi: minimum number of pixels, NPma: maximum number of pixels, NPa: average number of pixels

Table 2. Statistics on the estimation accuracy of the SMC using different methods (First scenario with corresponding scale parameter: Sc1-SP, Second scenario with corresponding scale parameter: Sc2-SP )

| Methods | Accuracy |
|---|---|

|          | RMSE (vol%) | Bias (vol%) | $R^2$ |
|----------|-------------|-------------|-------|
| S1S2     | 5.61        | -1.68       | 0.70  |
| SMAP     | 4.89        | -0.14       | 0.89  |
| Sc1-64   | 13.59       | -7.11       | 0.48  |
| Sc2-64   | 4.85        | -0.11       | 0.92  |
| Sc1-128  | 13.53       | -6.66       | 0.51  |
| Sc2-128  | 4.85        | -0.13       | 0.93  |
| Sc1-256  | 12.17       | -7.17       | 0.55  |
| Sc2-256  | 4.864       | -0.16       | 0.90  |
| Sc1-512  | 8.58        | -4.97       | 0.63  |
| Sc2-512  | 4.826       | -0.34       | 0.95  |
| Sc1-1024 | 7.88        | -4.68       | 0.68  |
| Sc2-1024 | 4.923       | -0.25       | 0.87  |
| Sc1-2048 | 11.82       | -8.06       | 0.59  |
| Sc2-2048 | 5.332       | -1.41       | 0.79  |

In Figure 5, the box and whisker plot has been drawn for different scenarios. It enables us to study the distributional characteristics of different approaches. As can be seen, there is a shorter box plot for the SMAP soil moisture map and second scenario (including SMAP data) than extracted soil moisture from coupled Sentinel-1 and Sentinel-2 and the first scenario (not including SMAP data). This indicates the smaller variation of soil moisture values in the SMAP soil moisture map. The more similar pattern (e.g., mean value) of the in-situ soil moisture box plot and SMAP data also confirm the accuracy assessment measures. It is also clear that the second scenario presents more homogenous behavior for different map scales than the first scenario.

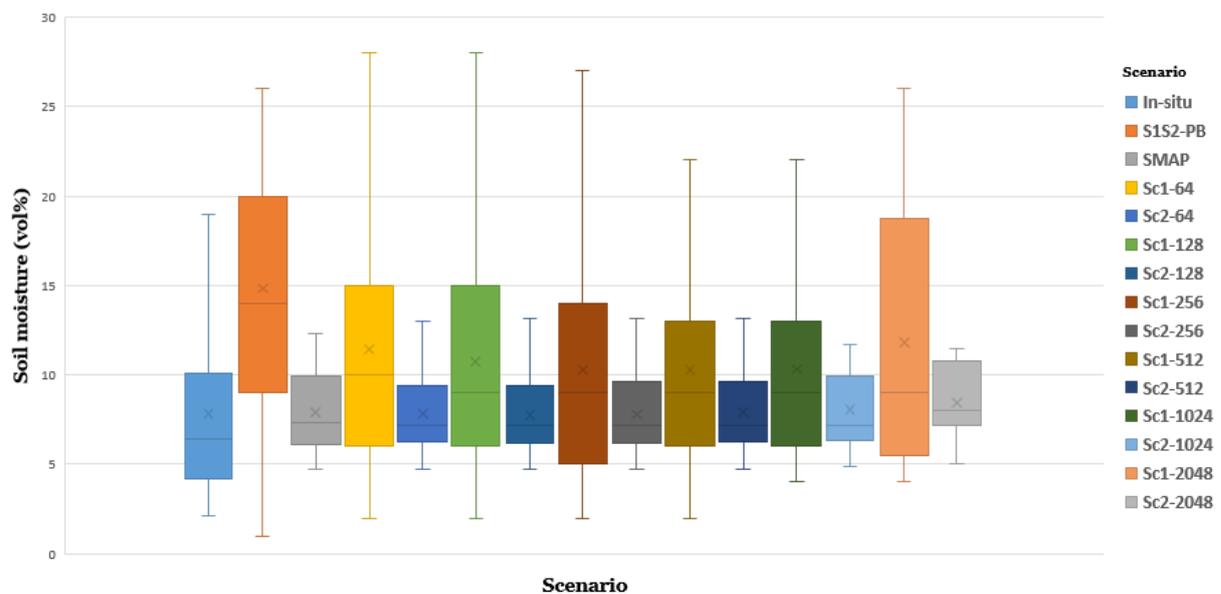

Figure 5. The box and whisker plot and the distributional characteristics of different

scenarios.

## 4. Conclusion

In this paper, we have presented the synergetic use of SMAP and Sentinel 1 & 2 data for preparing multi-scale soil moisture maps through the OBIA approach. Regarding the failure of the radar device of the SMAP mission and the impossibility of having a large-scale soil moisture map, there is this potential to exploits radar and optic Sentinel data to overcome this issue. The results of this study indicate that we can exploit the high accuracy and high spatial resolution of SMAP and Sentinel data respectively to produce a multi-scale soil moisture map. Regarding our previous work, we have obtained satisfactory results demonstrating that utilizing SMAP data could lead to more accurate results compared to employing only Sentinel data. The results demonstrated that exploiting the segment-based approach in the synergetic use of Sentinel-1, Sentinel-2 and SAMP data improved the accuracy of soil moisture estimation by 25% (as measured by $R^2$) compared to the pixel-based approach. There are two major challenges in this approach and it is recommended that further research should be undertaken in the following areas: The first is creating image objects related to soil moisture variable and the second is calculating soil moisture values for each image object.

**Acknowledgments**

The authors would like to thank Claudia Notarnicola and Felix Greifeneder for their support. The authors would also like to thank the Ministry of Science, Research, and Technology of Iran, and also the Institute for Earth Observation of Eurac Research (Italy) for the support of this paper. The authors thank the three anonymous reviewers for their very constructive remarks.

**References**

Amazirh, A., Merlin, O., Er-Raki, S., Gao, Q., Rivalland, V., Malbeteau, Y., Khabba, S., and Escorihuela, M. J. 2018. "Retrieving surface soil moisture at high spatio-temporal resolution from a synergy between Sentinel-1 radar and Landsat thermal data: A study case over bare soil." *Remote Sensing of Environment* 211, 321–337. DOI: 10.1016/j.rse.2018.04.013

Attarzadeh, R., and Amini, J. 2019a. "Investigating the Possibility of Preparing Small Scale Soil Moisture Map from Coupled SENTINEL-1 and SENTINEL-2 Data." *The International Archives of Photogrammetry, Remote Sensing and Spatial Information Sciences* 42, 127–129. DOI: 10.5194/isprs-archives-XLII-4-W18-127-2019

Attarzadeh, R., and Amini, J. 2019b. "Towards an object-based multi-scale soil moisture product using coupled Sentinel-1 and Sentinel-2 data." *Remote Sensing Letters* 10(7), 619–628. DOI: 10.1080/2150704X.2019.1590872

Attarzadeh, R., Amini, J., Notarnicola, C., and Greifeneder, F. 2018. "Synergetic use of Sentinel-1 and Sentinel-2 data for soil moisture mapping at plot scale." *Remote Sensing* 10(8), 1285. DOI: 10.3390/rs10081285

Bai, J., Cui, Q., Zhang, W., and Meng, L. 2019. "An approach for downscaling SMAP soil moisture by combining Sentinel-1 SAR and MODIS data." *Remote Sensing* 11(23), 2736. DOI: 10.3390/rs11232736

Bauer-Marschallinger, B., Freeman, V., Cao, S., Paulik, C., Schaufler, S., Stachl, T., Modanesi, S., Massari, C., Ciabatta, L., Brocca, L., and Wagner, W. 2018. "Toward global soil moisture monitoring with Sentinel-1: Harnessing assets and overcoming obstacles." *IEEE Transactions on Geoscience and Remote Sensing* 57(1), 520–539. DOI: 10.1109/TGRS.2018.2858004

Bazzi, H., Baghdadi, N., Ienco, D., El Hajj, M., Zribi, M., Belhouchette, H., Escorihuela, M. J., and Demarez, V. 2019. "Mapping Irrigated Areas Using Sentinel-1 Time Series in Catalonia, Spain". *Remote Sensing*, 11(15). doi:10.3390/rs11151836

Das, N. N., Entekhabi, D., Dunbar, R. S., Chaubell, M. J., Colliander, A., Yueh, S., Jagdhuber, T., Chen, F., Crow, W., O'Neill, P.E., Walker, J.P., Berg, A., Bosch, D.D., Caldwell, T., Coshm M.H., Collins, C.H., Lopez-Baeza, E., and Thibeault, M. 2019. "The SMAP and Copernicus Sentinel 1A/B microwave active-passive high resolution surface soil moisture product." *Remote Sensing of Environment* 233, 111380. DOI: 10.1016/j.rse.2019.111380

Das, N. N., Entekhabi, D., Dunbar, R. S., Colliander, A., Chen, F., Crow, W., Jackson, T.J., Breg, A., Bosch, D.D., Caldwell, T., Cosh, M.H., Collins, C.H., Lopez-Baeza, E., Moghaddam, M., Rowlandson, T., Starks, P.J., Thibeault, M., Walker, J.P., Wu, X., O'Neill, P.E., Yueh, S., and Njoku, E. G. 2018. "The SMAP mission combined active-passive soil moisture product at 9 km and 3 km spatial resolutions." *Remote Sensing of Environment*

211, 204–217. DOI: 10.1016/j.rse.2018.04.011

El Hajj, M., Baghdadi, N., Zribi, M., Rodríguez-Fernández, N., Wigneron, J. P., Al-Yaari, A., Al Bitar, A., Albergel, C., and Calvet, J. C. 2018. "Evaluation of SMOS, SMAP, ASCAT and Sentinel-1 soil moisture products at sites in Southwestern France." *Remote Sensing* 10(4), 569. DOI: 10.3390/rs10040569

He, L., Hong, Y., Wu, X., Ye, N., Walker, J. P., and Chen, X. 2018. "Investigation of SMAP active–passive downscaling algorithms using combined sentinel-1 SAR and SMAP radiometer data." *IEEE Transactions on Geoscience and Remote Sensing* 56(8), 4906–4918. DOI: 10.1109/TGRS.2018.2842153

Jackson, T. J., Cosh, M. H., Bindlish, R., Starks, P. J., Bosch, D. D., Seyfried, M., … Du, J. 2010. "Validation of Advanced Microwave Scanning Radiometer Soil Moisture Products". *IEEE Transactions on Geoscience and Remote Sensing*, 48(12), 4256–4272.

Jagdhuber, T., Baur, M., Akbar, R., Das, N. N., Link, M., He, L., and Entekhabi, D. 2019. "Estimation of active-passive microwave covariation using SMAP and Sentinel-1 data." *Remote Sensing of Environment* 225, 458–468. DOI: 10.1016/j.rse.2019.03.021

Karthikeyan, L., Pan, M., Wanders, N., Kumar, D. N., and Wood, E. F. 2017. "Four decades of microwave satellite soil moisture observations: Part 1. A review of retrieval algorithms." *Advances in Water Resources* 109, 106–120. DOI: 10.1016/j.advwatres.2017.09.006

Koster, R. D., Dirmeyer, P. A., Guo, Z., Bonan, G., Chan, E., Cox, P., Gordon, C. T., Kanae, S., Kowalczyk, E., Lawrence, D., Liu, P., Lu, C. H., Malyshev, S., McAvaney, B., Mitchell, K., Mocko, D., Oki, T., Oleson, K., Pitman, A., Sud, Y. C., Taylor, C. M., Verseghy, D., Vasic, R., Xue, Y., Yamada, T., and GLACE Team. 2004. "Regions of strong coupling between soil moisture and precipitation". *Science*, 305(5687), 1138–1140. https://doi.org/10.1126/science.1100217

Leone, D. 2015. "NASA Focused on Sentinel as Replacement for SMAP Radar." Space News. Available online at: http://spacenews.com/nasa-focused-on-sentinel-as-replacementfor-smap-radar (Accessed on 02 Jan 2018).

Motte, E., et al. 2016. GLORI: "A GNSS-R Dual Polarization Airborne Instrument for Land Surface Monitoring". *Sensors*, 16(5).

Sabaghy, S., Walker, J. P., Renzullo, L. J., and Jackson, T. J. 2018. "Spatially enhanced passive microwave derived soil moisture: Capabilities and opportunities." *Remote Sensing of Environment* 209, 551–580. DOI: 10.1016/j.rse.2018.02.065

Santi, E., Paloscia, S., Pettinato, S., Brocca, L., Ciabatta, L., and Entekhabi, D. 2018. "On the synergy of SMAP, AMSR2 AND SENTINEL-1 for retrieving soil moisture." *International Journal of Applied Earth Observation and Geoinformation* 65, 114–123. DOI: 10.1016/j.jag.2017.10.010

Wagner, W., et al. 2013. "The ASCAT Soil Moisture Product: A Review of its Specifications,

Validation Results, and Emerging Applications". *Meteorologische Zeitschrift*, 22(1), 5–33.